\documentclass[review]{elsarticle}

\journal{Computer Physics Communications}

\usepackage{lineno,hyperref}
\usepackage{bm,amsmath,amssymb}
\usepackage{graphicx}

\newcommand{\tc}{T_{\rm c}}
\newcommand{\nef}{N(0)}

\newcommand{\omegalog}{\omega_{\rm log}}

\begin{document}

\begin{frontmatter}

\title{Efficient method to calculate the electron-phonon coupling constant and superconducting transition temperature}

\author[mymainaddress,mysecondaryaddress]{Takashi Koretsune}
\address[mymainaddress]{RIKEN, Center for Emergent Matter Science, 2-1 Hirosawa, Wako, Saitama 351-0198, Japan}
\address[mysecondaryaddress]{JST, PRESTO, 4-1-8 Honcho, Kawaguchi, Saitama 332-0012, Japan}
\cortext[mycorrespondingauthor]{Corresponding author}
\ead{takashi.koretsune@riken.jp}

\author[mymainaddress,mythirdaddress]{Ryotaro Arita}
\address[mythirdaddress]{ERATO Isobe Degenerate $\pi$-Integration Project, Tohoku University, Aoba-ku, Sendai 980-8578, Japan}

\begin{abstract}
We show an efficient way to compute the electron-phonon coupling constant, $\lambda$, and the superconducting transition temperature, $\tc$ from first-principles calculations.
This approach gives rapid convergence of $\tc$ with respect to the size of the $\bm k$-point mesh,
and is seamlessly connected to the formulation used in large molecular systems such as alkali fullerides where momentum dependence can be neglected.
Since the phonon and electron-phonon calculations are time consuming particularly in complicated systems,
the present approach will strongly reduce the computational cost,
which facilitates high-throughput superconducting material design.

\end{abstract}

\begin{keyword}
electron-phonon coupling\sep superconductivity\sep k-space integration
\end{keyword}

\end{frontmatter}

%\linenumbers

\section{Introduction}
Precise prediction of the superconducting transition temperature, $\tc$, and designing high-temperature superconducting materials are one of the ultimate goals for material design.
Particularly for phonon-mediated superconductivity,
several recent studies based on superconducting density functional theory (SCDFT)\cite{Luders2005,Marques2005} and {\it ab-initio} Eliashberg approach\cite{sano2016} show the quantitative agreement of experimental $\tc$ from fully {\it ab-initio} calculations.
Furthermore, recent studies\cite{Li2014,Duan2014} successfully predicted the structure of sulfur hydrides at high pressures and its superconductivity preceding the experimental discovery\cite{Drozdov2015}.
Thanks to these studies, exploring superconducting materials has become a more realistic and promising issue.
On the other hand, {\it ab-initio} calculation of $\tc$,
particularly, calculating the electron-phonon couplings and obtaining their convergence are still time-consuming tasks.
Thus, a high-throughput method to evaluate the electron-phonon couplings is required to study a wide range of materials.

The key quantity that represents the strength of the electron-phonon couplings is called $\lambda$, which is basically obtained by averaging the electron-phonon couplings over the Fermi surface.
It is known that this average converges very slowly with respect to the size of $\bm k$-point mesh.
In this paper, we show simple but efficient method to evaluate this average over the Fermi surface.
By applying this method to Pb and Nb, we demonstrate that the convergence of $\lambda$ and $\tc$ with respect to $\bm k$-point mesh size is improved compared to the conventional Gaussian smearing method.

\section{Method}
To evaluate $\tc$, the standard approach is to solve the Eliashberg equation by introducing the energy cutoff and the pseudo Coulomb potential, $\mu^*$\cite{allen1983}.
The powerful and convenient expressions of $\tc$ that approximate this Eliashberg equation are McMillan\cite{mcmillan1968} and Allen-Dynes\cite{allen1975} formulas.
In the Allen-Dynes formula, $\tc$ is given as
\begin{align}
	\tc = \frac{\omegalog}{1.2} \exp \left( - \frac{1.04(1+\lambda)}{\lambda - \mu^*(1 + 0.62 \lambda)} \right).
\end{align}
Here, $\mu^*$ is a pseudo Coulomb potential and $\lambda$ and $\omegalog$ are defined as
\begin{align}
	\lambda &= 2 \int d \omega \frac{\alpha^2 F(\omega)}{\omega},\label{eq:lambda}\\
	\ln \omegalog &= \frac{2}{\lambda} \int d \omega \frac{\alpha^2 F(\omega)}{\omega} \ln(\omega) \label{eq:omegalog}
\end{align}
with
\begin{align}
	\alpha^2 F(\omega) = \frac{1}{\nef} 
	%\sum_{n \bm k, m \bm q, \nu} | g_{n \bm k, m \bm k + \bm q}^{\nu} |^2 \delta(\epsilon_{n \bm k} - \ef) \delta(\epsilon_{m \bm k+ \bm q} - \ef)
	\sum_{n \bm k, m \bm q, \nu} | g_{n \bm k, m \bm k + \bm q}^{\nu} |^2 \delta(\xi_{n \bm k}) \delta(\xi_{m \bm k+ \bm q})
	\delta(\omega - \omega_{\nu \bm q}).
	\label{eq:alpha2F}
\end{align}
Here, $\nef = \sum_{n \bm k}\delta(\xi_{n \bm k})$ is the density of states at the Fermi level, $\xi_{n \bm k}$ is a one-particle band energy with respect to the Fermi level at band index, $n$ and wave vector, $\bm k$, $\omega_{\nu \bm q}$ is the phonon frequency at phonon mode $\nu$ and wave vector, $\bm q$,  and $g_{n \bm k,m \bm k + \bm q}^\nu$ is the electron-phonon coupling.
The simple approach to evaluate Eq.\ \eqref{eq:alpha2F} is to take a discrete summation on finite $\bm k$- and $\bm q$-point meshes by replacing two $\delta$ functions with smearing functions with an appropriate smearing width.
Thus, to get the convergence, one needs to increase the size of $\bm k$- and $\bm q$-point meshes with decreasing the smearing width.
However, it is known that the convergence of this approach is extremely slow.
Even the tetrahedron method\cite{allen1983,savrasov1996} needs a sufficient $\bm k$- and $\bm q$-point meshes.
To overcome the problem, the interpolation of electron-phonon coupling using Wannier function technique has been developed\cite{Giustino2007,EPW2016}.

To circumvent the problem for the summation of the two $\delta$ functions, we transform Eqs.\ \eqref{eq:lambda}-\eqref{eq:alpha2F} as follows:
\begin{align}
	\lambda &= \frac{2}{\nef} \sum_{n \bm k, m \bm q, \nu} \frac{| g_{n \bm k, m \bm k + \bm q}^{\nu} |^2}{\omega_{\nu \bm q}}
	\delta(\xi_{n \bm k}) \delta(\xi_{m \bm k+ \bm q})\nonumber\\
	&= 2 \nef \left\langle \sum_{\nu} \frac{| g_{n \bm k, m \bm k + \bm q}^{\nu} |^2}{\omega_{\nu \bm q}} \right\rangle,
	\label{eq:lambda_weight}\\
	\ln \omegalog &= \frac{2 \nef}{\lambda} 
	\left\langle \sum_{\nu} \frac{ | g_{n \bm k, m \bm k + \bm q}^{\nu} |^2 \ln \omega_{\nu \bm q} }{\omega_{\nu \bm q}} \right\rangle.
	\label{eq:omegalog_weight}
\end{align}
Here, $\left\langle \cdots \right\rangle$ is a weighted average defined as
\begin{align}
	\left\langle O_{n m \bm k \bm q} \right\rangle &= \sum_{n \bm k, m \bm q} O_{n m \bm k \bm q} P_{nm\bm k \bm q},\\
	P_{nm\bm k \bm q} &= \frac{\delta(\xi_{n \bm k}) \delta(\xi_{m \bm k + \bm q})} {\sum_{n' \bm k', m' \bm q'}{\delta(\xi_{n' \bm k'}) \delta(\xi_{m' \bm k' + \bm q'})}}.
\end{align}
Here the probability $P_{nm \bm k\bm q}$ satisfies $\sum_{nm\bm k\bm q} P_{nm \bm k \bm q} = 1$ even when we replace the $\delta$-functions with other smearing functions.
Thus, if $O_{nm\bm k\bm q}$ does not depend much on $\bm k$ and $\bm q$, we do not need dense $\bm k$ and $\bm q$ meshes.
In the practical calculation, we estimate $\nef$ using the tetrahedron method with a dense $\bm k$-point mesh while for $P_{nm\bm k\bm q}$ we use the Gaussian smearing functions instead of the $\delta$ functions.
It should be noted that this method can be also regarded as a rescaling of $\lambda$ by $N(0)^2/ \sum_{n \bm k, m \bm q}{\delta(\xi_{n \bm k}) \delta(\xi_{m \bm k + \bm q})}$.
Here, the denominator and numerator should be the same in principle, while only for the denominator, we employ a coarse mesh used in the electron-phonon coupling calculation.
Therefore, this method can be applicable not only for the Gaussian smearing method, but also for the tetrahedron method or other smearing methods and can be implemented very easily.
We implemented this method based on the quantum-{\sc ESPRESSO} package\cite{espresso}.
The code is distributed through the website\cite{lambda} and is available under the terms of the GNU General Public License.

\section{Result}
In the following, we illustrate the comparison of the weighted-average method and the conventional Gaussian smearing method for the case of Pb and Nb.
We obtain the electronic structures, phonon properties and electron-phonon couplings using the density functional theory (DFT) and the density functional perturbation theory (DFPT) as implemented in the quantum-{\sc ESPRESSO} code\cite{espresso}.
We employ the local density approximation (LDA)\cite{pz} and a norm-conserving pseudopotential with the cutoff energy of 90 Ry for Pb and the generalized gradient approximation\cite{pbe} and an ultrasoft pseudopotential\cite{vanderbilt1990} with the cutoff energies of 30 Ry for wavefunctions and 300 Ry for charge densities for Nb.

\subsection{Density of states at the Fermi level}
Before discussing the electron-phonon couplings, we first show the convergence of the density of states at the Fermi level, $N(0) = \sum_{n \bm k} \delta(\xi_{n\bm k})$, using the Gaussian smearing method and the tetrahedron method.
This roughly illustrates how the electron-phonon couplings converge since when we neglect the momentum and band dependence of the electron-phonon couplings and phonon frequency, $i.e.$, $g_{n \bm k, m \bm k + \bm q}^{\nu} =$ const, $\omega_{\nu\bm q} =$ const, using Eq.\ \eqref{eq:lambda_weight}, we obtain $N(0) \lambda \propto \left(\sum_{n \bm k} \delta(\xi_{n\bm k}) \right)^2$.

Figure 1 shows $N(0)$ for (a) Pb and (b) Nb as a function of the Gaussian smearing width, $\delta$, for several $\bm k$-point meshes and
insets show $N(0)$ as a function of the $\bm k$-point mesh with $\delta = 0.02$ and $0.05$ Ry and the tetrahedron method.
Using the Gaussian smearing method, we should check the convergence by decreasing the smearing width and increasing the $\bm k$-point mesh.
For Pb, to get the convergence at $\delta = 0.05$ and $0.02$, we need $\sim 16^3$ and $\sim 30^3$ $\bm k$ points, respectively, as shown in the inset of Fig. 1(a).
However, the converged value at each $\delta$ does not much depend on $\delta$ as typically seen in $32^3 \bm k$-point calculation.
In fact, the converged value of $N(0)$ obtained by tetrahedron method is $N(0) = 0.49$, which is consistent with the value at $\delta = 0.01$ and $32^3 \bm k$-point calculation.
For Nb, on the other hand, the convergence at $\delta = 0.05$ and $0.02$ can be obtained for smaller number of $\bm k$ points of $\sim 16^3$ and $\sim 20^3$, respectively.
In this case, however, the converged value at each $\delta$ highly depends on $\delta$.
In fact, $N(0)$ for $32^3 \bm k$-point calculation is still increasing at the smallest $\delta$ and from this calculation, it is difficult to extract the converged value of $N(0) = 1.49$ obtained by the tetrahedron method.
It should be noted that even in the tetrahedron method, the convergence is achieved for $30^3 \sim 40^3$ $\bm k$ points for Nb.

\begin{figure}
	\label{fig:dos_all}
	\begin{center}
	\includegraphics[width=0.45\textwidth]{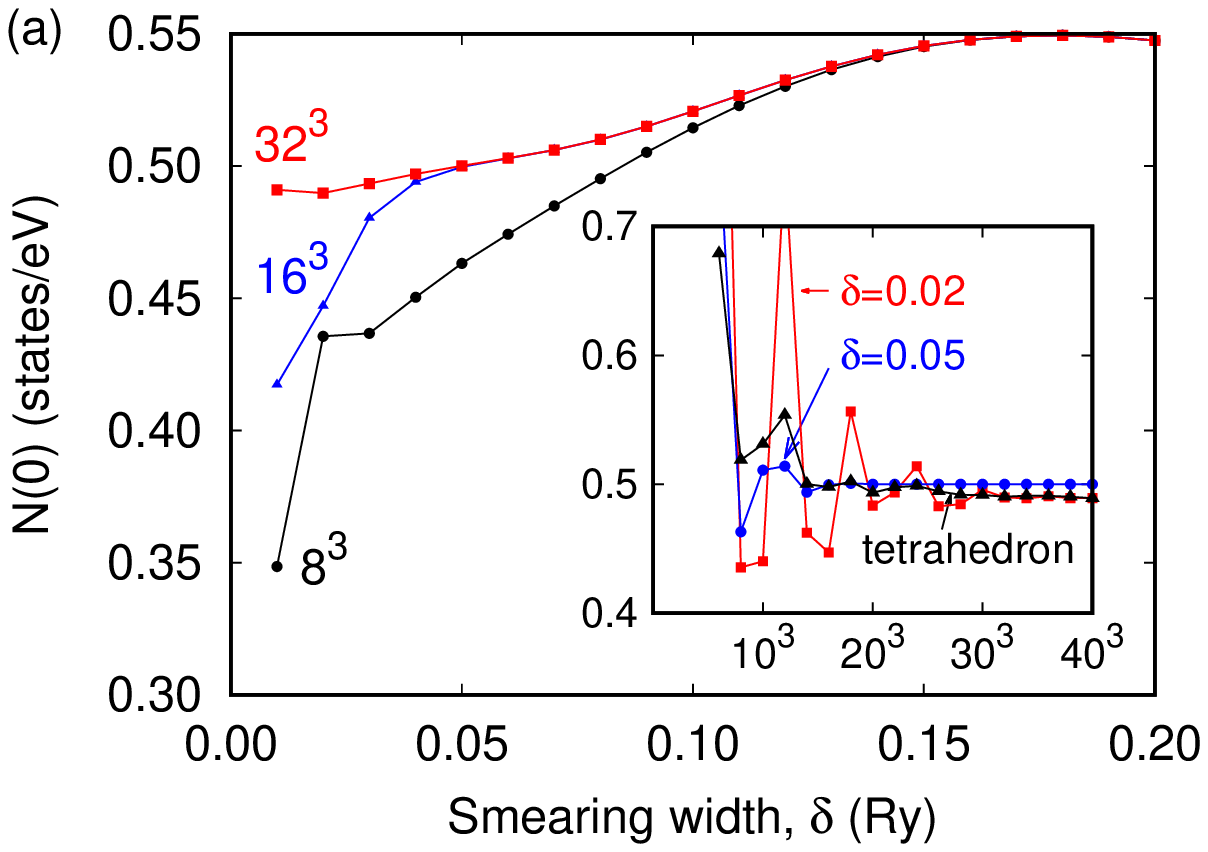}
	\includegraphics[width=0.45\textwidth]{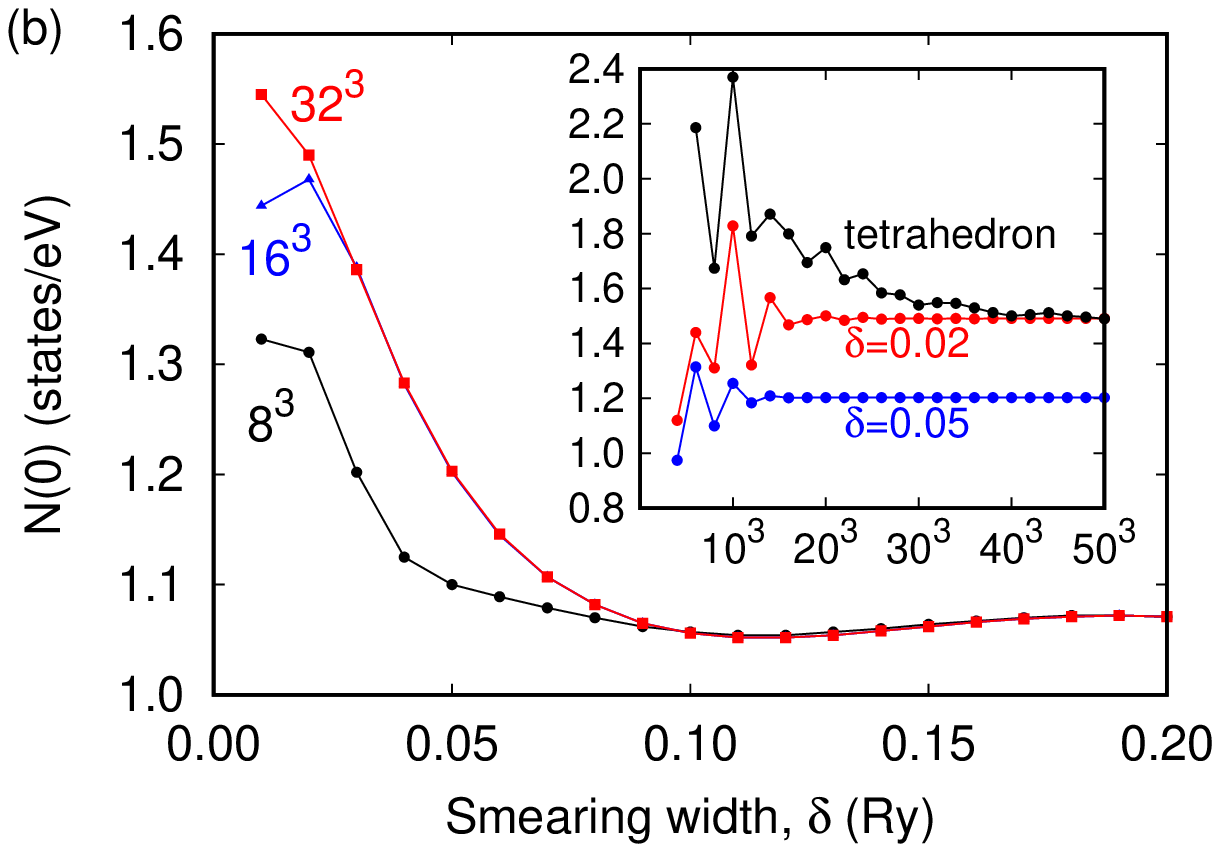}
	\vspace{1cm}
	\caption{
		Densities of states at the Fermi level, $N(0)$, for (a) Pb and (b) Nb as a function of the Gaussian smearing width, $\delta$, with $\bm k$-point mesh of $8\times8\times8$, $16\times16\times16$, and $32\times32\times32$. Insets show $N(0)$ as a function of $\bm k$-point mesh with the smearing width of 0.02 and 0.05 Ry and using the tetrahedron method.
	}
	\end{center}
\end{figure}

\subsection{Electron-phonon couplings}
In the calculation of the electron-phonon coupling constant, $\lambda$, we need to consider the convergence with respect to both $\bm k$- and $\bm q$-point meshes as explained above.
In practical calculation, $\bm k$-point mesh is more important because $\bm k$-point mesh also determines the accuracy of the phonon frequency, $\omega_{\nu \bm q}$ and electron-phonon couplings, $g_{n \bm k, m \bm k + \bm q}^{\nu}$.
In this calculation, for simplicity, we discuss the convergence using the same mesh size, $n \times n \times n$ with $n = 8, 12, 16,$ and $20$, for $\bm k$ and $\bm q$ points.
That is, we compute the electronic structures using the uniform $n \times n \times n$ $\bm k$-point mesh with Gaussian smearing width of 0.025 Ry.
Then, we obtain the dynamical matrix and electron-phonon couplings with $n \times n \times n$ $\bm k$ and $\bm q$-point meshes.

\begin{figure}
	\includegraphics[width=0.5\textwidth]{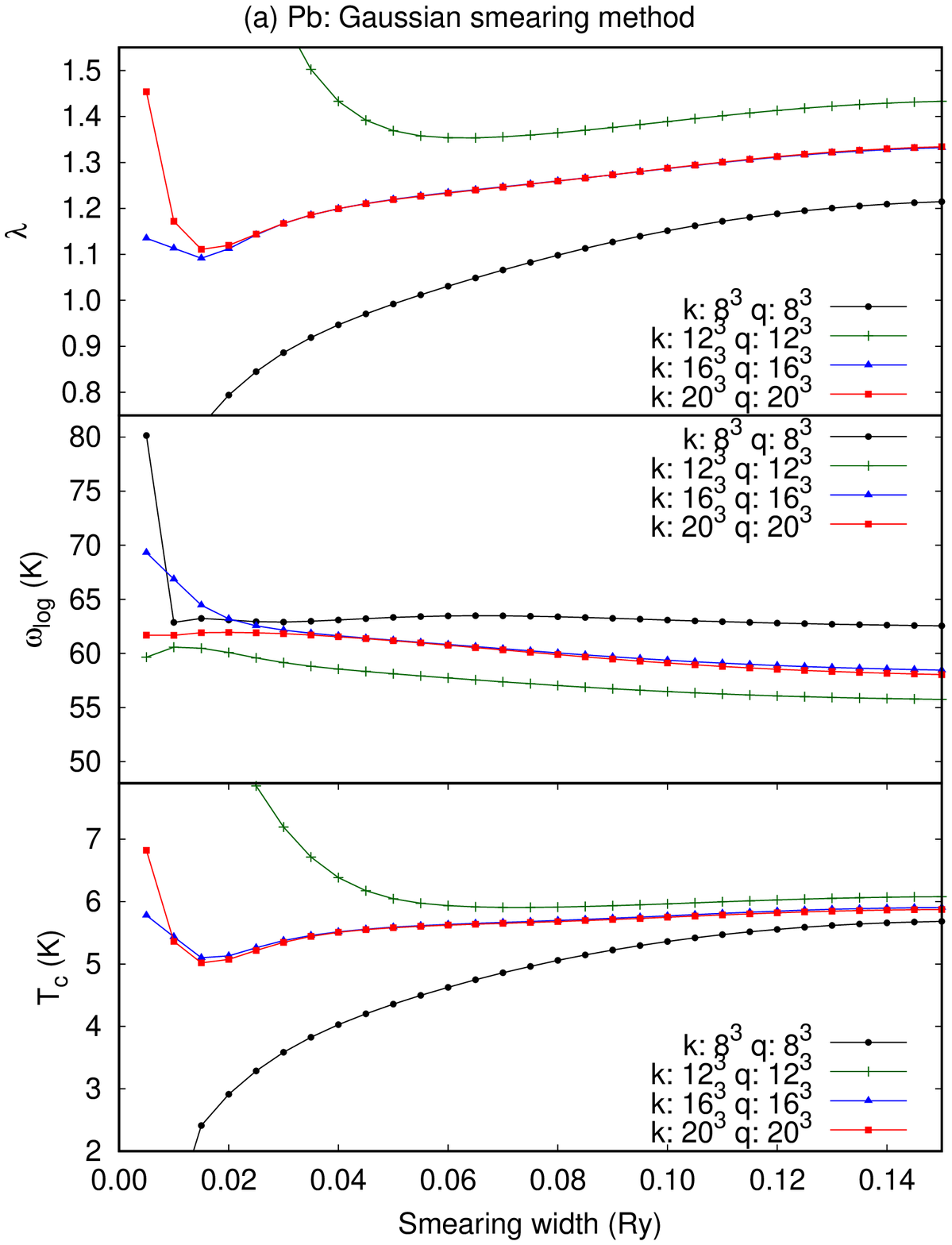}
	\includegraphics[width=0.5\textwidth]{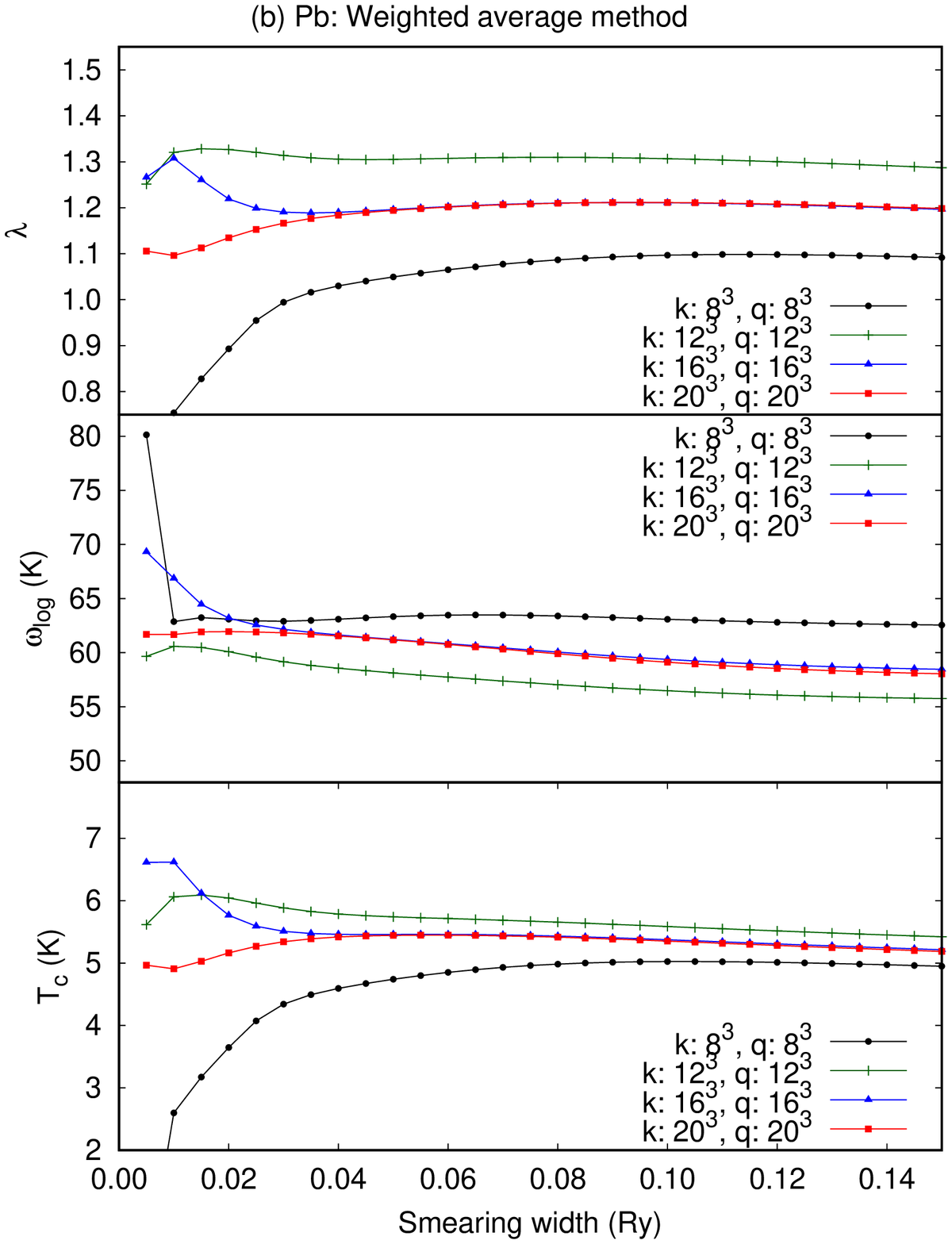}
	\caption{
		Electron-phonon coupling constant, $\lambda$, logarithmic average of phonon frequency, $\omega_{\rm log}$, and transition temperature, $T_{\rm c}$ at $\mu^*=0.10$ for Pb,
		calculated using (a) the conventional Gaussian smearing method and (b) the weighted-average method.
	}
	\label{fig:smearing_width_dep_pb}
\end{figure}

Figure \ref{fig:smearing_width_dep_pb} shows the electron-phonon coupling constant, $\lambda$, logarithmic average of phonon frequency, $\omega_{\rm log}$, and transition temperature, $T_{\rm c}$ at $\mu^*=0.10$ for Pb.
In the conventional smearing method, $\bm k$-point mesh dependence of $\lambda$ and $T_{\rm c}$ is large particularly for $\delta < 0.04$ Ry and it is difficult to get the convergence, which is consistent with the discussion for $N(0)$.
In fact, Ref. \cite{EPW2016} discussed the convergence of $\lambda$ in more dense meshes using the EPW code and more than $60^3$ $\bm k$ points are needed for $\delta < 0.005$ Ry to obtain the converged value of $\lambda = 1.1 \sim 1.2$.
In the weighted-average method, on the other hand, $\lambda$, $\omega_{\rm log}$ and $T_{\rm c}$ do not depend much on $\delta$ as shown in Fig.\ \ref{fig:smearing_width_dep_pb}(b).
The obtained values of $\lambda = 1.1 \sim 1.3$, $\omega_{\rm log} = 60 \sim 65$ K and $T_{\rm c} = 5 \sim 6$ K are consistent with the EPW calculation with dense $\bm k$ and $\bm q$-point meshes.

\begin{figure}
	\includegraphics[width=0.5\textwidth]{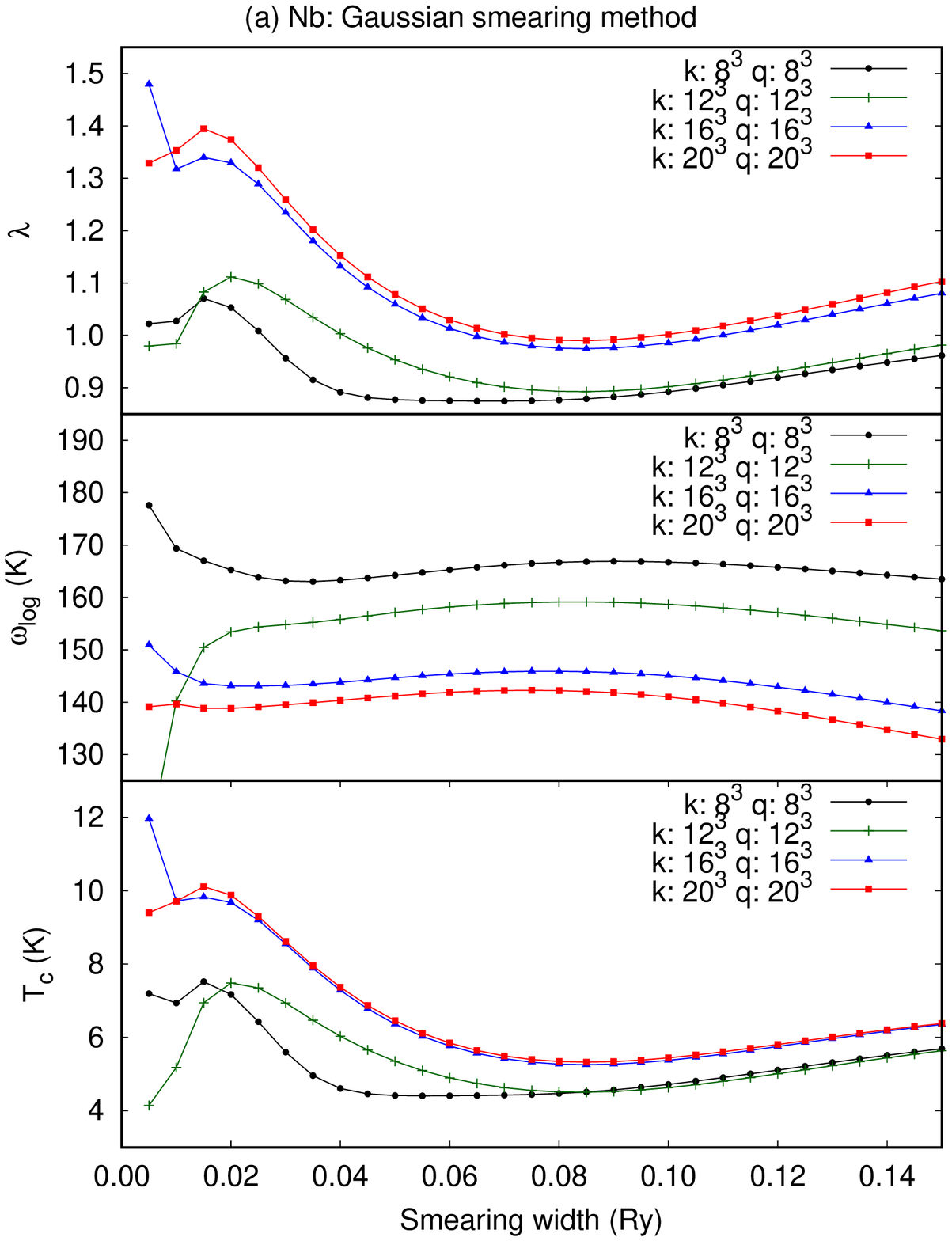}
	\includegraphics[width=0.5\textwidth]{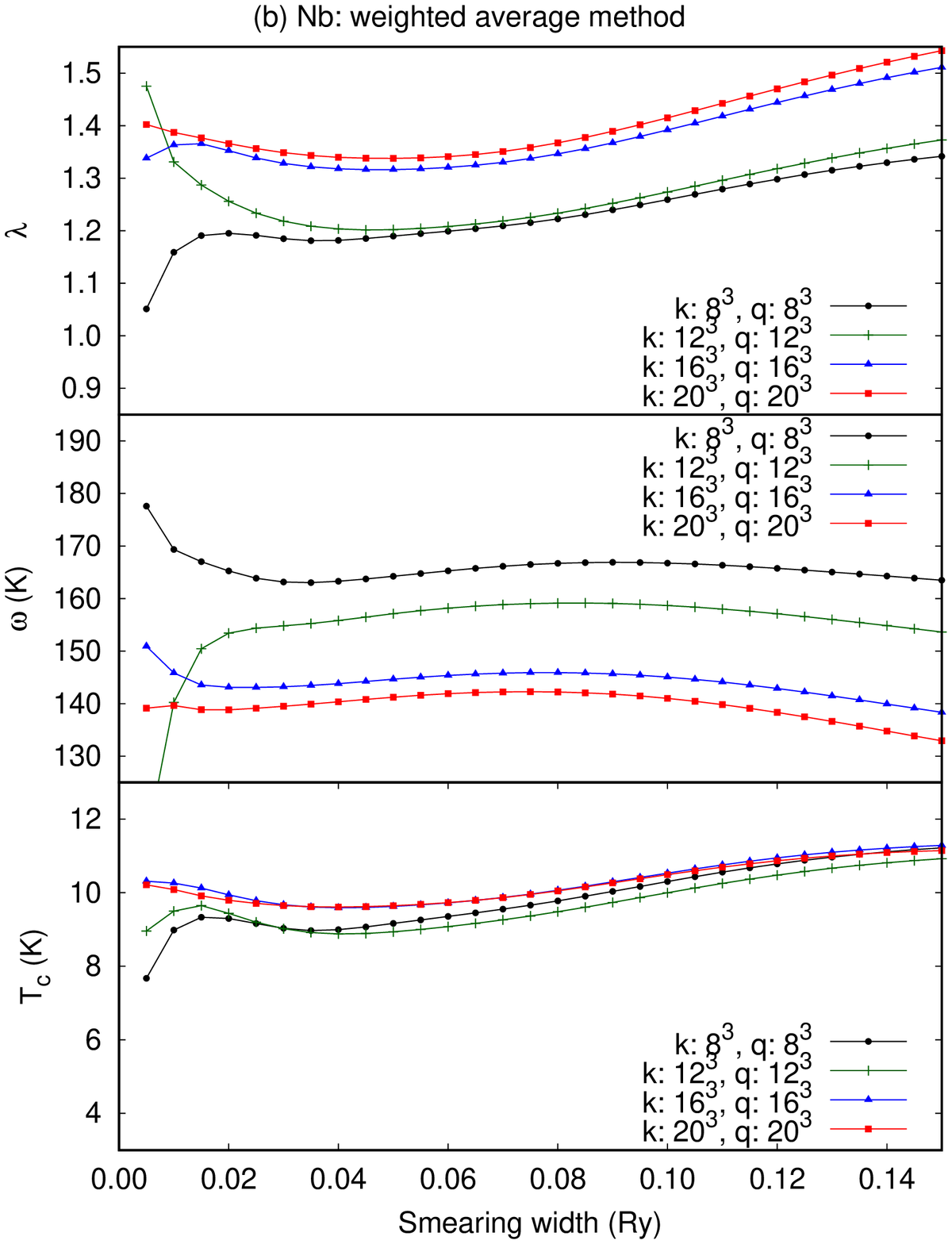}
	\caption{
		Electron-phonon coupling constant, $\lambda$, logarithmic average of phonon frequency, $\omega_{\rm log}$, and transition temperature, $T_{\rm c}$ at $\mu^*=0.20$ for Nb,
		calculated using (a) the conventional Gaussian smearing method and (b) the weighted-average method.
	}
	\label{fig:smearing_width_dep_nb}
\end{figure}

In the case of Nb, the $\delta$ dependence in the conventional smearing method is more severe as shown in Fig. \ref{fig:smearing_width_dep_nb}.
$\lambda$ and $\tc$ show similar $\delta$ dependence with $N(0)$ and it is difficult to evaluate $\lambda$ and $T_{\rm c}$.
On the other hand, using the weighted-average method, $\delta$ dependence is small and we can estimate $\lambda = 1.3 \sim 1.4$, $\omega_{\rm log} = 140 \sim 160$ K and $T_{\rm c} \sim 10$ K for $\mu^* = 0.20$.
One can see a large $\bm k$-point size dependence in $\lambda$ and $\omegalog$ even at large $\delta$ while this almost cancels out in $\tc$.
This is often the case for the electron-phonon coupling calculations and can be explained by low phonon-frequency modes.
Namely, it is difficult to get the accurate $\omega_{\nu \bm q}$ and $g^{\nu}_{\bm k,\bm k+\bm q}$ for low phonon frequencies, while low-frequency modes can give large contributions to $\lambda$ and $\omegalog$ due to $\omega_{\nu \bm q}^{-1}$ dependence in Eqs.\ \eqref{eq:lambda} and \eqref{eq:omegalog}.
However, these contributions have opposite effects on $\tc$ and as a result, $\tc$ is less sensitive to the accuracy of low phonon-frequency modes.

The important result in the weighted-average method is that $\delta$ dependence is significantly suppressed compared to the conventional Gaussian smearing method in both Pb and Nb cases.
For example, by changing $\delta$ from 0.15 to 0.02, $\tc$ typically changes only about 10\%.
This indicates that the momentum dependence of $|g^{\nu}_{\bm k,\bm k+\bm q}|^2$ is weak and the difficulty in summation of Eq.\ \eqref{eq:alpha2F} in the conventional method comes mainly from two $\delta$ functions.
Particularly for $\lambda$ and $\omegalog$, $\bm k$-point size dependence is comparable to the $\delta$ dependence suggesting that the accuracy of $\omega_{\nu \bm q}$ and $g^{\nu}_{\bm k,\bm k+\bm q}$ can be more important issue than the mesh size of the summation in Eqs.\ \eqref{eq:lambda_weight} and \eqref{eq:omegalog_weight}.
Using this small $\delta$ dependence, one interesting application is a systematic rough estimate of electron-phonon couplings for a wide range of materials with a large smearing width and small number of $\bm k$ and $\bm q$-points for a screening purpose.

Another advantage of this method is that when we can neglect the momentum and band-index dependence in $g_{n \bm k, m \bm k + \bm q}^{\nu}$ and $\omega_{\nu \bm q}$, the result becomes accurate.
In fact, Eq.\ \eqref{eq:lambda_weight} falls into the formula used in large molecular systems such as alkali fullerides where momentum dependence can be neglected\cite{gunnarsson1997}.
Thus, we can choose an appropriate $\bm k$- and $\bm q$-point mesh size depending on the momentum dependence of $g_{n \bm k, m \bm k + \bm q}^{\nu}$ and $\omega_{\nu \bm q}$ and can estimate $\tc$ even for large systems with practical computational time.

\section{Summary}
We presented an efficient method to calculate $\lambda$ and $\tc$ and demonstrated that this method significantly improves the convergence with respect to the $\bm k$- and $\bm q$-point mesh size.
We can use this method not only for accurate estimate but also for rough estimate of $\lambda$ and $\tc$ with coarse $k$ and $q$ meshes where even the tetrahedron method does not converge.
This method can be understood as a generalization of the formula used in large molecular systems where momentum dependence can be neglected and can be applicable even for large systems.
The method can be easily implemented and the example implementation is distributed through the website.

\section{Acknowledgment}
This work is financially supported by JST, PRESTO and JSPS KAKENHI Grant Numbers JP15H03696, JP16H00924.
\section*{References}

%\bibliography{manuscript}

\end{document}